\documentstyle[preprint,aps]{revtex}
\tighten
\draft
\widetext
\preprint{YITP-00-36}
\bigskip
\bigskip
\begin{document}
\title{Gravity in Randall-Sundrum Brane World Revisited}
\medskip
\author{Zurab Kakushadze\footnote{E-mail: 
zurab@insti.physics.sunysb.edu}}
\bigskip
\address{C.N. Yang Institute for Theoretical Physics\\ 
State University of New York, Stony Brook, NY 11794}

\date{August 15, 2000}
\bigskip
\medskip
\maketitle

\begin{abstract} 
{}We point out some subtleties with gauge fixings (which sometimes include
the so-called ``brane bending'' effects) typically used to 
compute the graviton propagator on the Randall-Sundrum brane. In particular,
the brane, which has non-vanishing tension, explicitly breaks some part of 
the diffeomorphisms, so that there are subtleties arising in going to, say, 
the axial gauge or the harmonic gauge
in the presence of (non-conformal) matter localized on the brane. 
We therefore compute the graviton 
propagator in the gauge where only the graviphoton fluctuations are set to 
zero
(the diffeomorphisms necessary for this gauge fixing are intact), but the 
graviscalar component is untouched. We point out that in the Gaussian
normal coordinates (where the graviscalar component vanishes on the brane)
the graviton propagator blows up in the ultra-violet near the brane.
In fact, the allowed gauge transformations, which do not lead to such
ultra-violet behavior of the graviton propagator, are such that 
the coupling of the graviscalar to the brane matter cannot be gauged away
in the ultra-violet. Because of this,
at the quantum level, where
we expect various additional terms to be generated in the brane world-volume
action including those involving the graviscalar, 
fine-tuning (which is independent of that for the brane cosmological 
constant) is generically required to preserve consistent 
coupling between bulk gravity and brane matter. We also reiterate that in such
warped backgrounds higher curvature terms in the bulk are generically
expected to delocalize gravity.  
\end{abstract}

\section{The Model}

{}In the Brane World scenario the Standard Model gauge and matter fields
are assumed to be localized on  
branes (or an intersection thereof), while gravity lives in a larger
dimensional bulk of space-time 
\cite{early,BK,polchi,witt,lyk,shif,TeV,dienes,3gen,anto,ST,BW}. The volume
of dimensions transverse to the branes is 
automatically finite if these dimensions are compact. On the other hand, 
the volume of the transverse dimensions
can be finite even if the latter are non-compact. In particular, this can be
achieved by using \cite{Gog} warped compactifications \cite{Visser} which
localize gravity on the brane. A concrete
realization of this idea was given in \cite{RS}.

{}In this paper we study gravitational interactions between matter sources
localized on the brane in the Randall-Sundrum model, whose action
is given by:
\begin{equation}\label{action}
 S=-f\int_\Sigma d^{D-1} x \sqrt{-{\widehat G}} +
 M_P^{D-2}
 \int d^D x \sqrt{-G} \left[R-\Lambda\right]~.
\end{equation}
For calculational convenience we will keep the number of space-time
dimensions $D$ unspecified.
In (\ref{action}) $M_P$ is the $D$-dimensional (reduced) Planck scale, 
while $\Lambda$ is the bulk vacuum energy density, which is assumed to be 
a negative constant. 
The $(D-1)$-dimensional hypersurface $\Sigma$, that is, the brane,
is the $z=0$ slice of the $D$-dimensional 
space-time, where $z\equiv x^D$. Next, 
\begin{equation}
 {\widehat G}_{\mu\nu}\equiv{\delta_\mu}^M {\delta_\nu}^N G_{MN}
 \Big|_{z=0}~,
\end{equation} 
where the capital Latin indices $M,N,\dots=1,\dots,D$, while the Greek
indices $\mu,\nu,\dots=1,\dots,(D-1)$. The quantity $f$ is
the brane tension\footnote{There might be additional bulk fields other than 
gravity. We will assume that such fields have vanishing expectation values.
Also, there might be various fields localized on the brane. If the brane 
world-volume theory is not conformal, then, as was pointed out in 
\cite{DGP,DG},
quantum loops of these fields are generically expected to induce, among 
other terms, the $(D-1)$
dimensional Einstein-Hilbert term on the brane. We will discuss the 
consequences of this at the end of the paper. For now, however, we will
ignore any such terms and stick to the original action (\ref{action}).}.

{}The equations of motion following from the action (\ref{action}) read:
\begin{equation}\label{EinsteinEOM}
 R_{MN}-{1\over 2}G_{MN} R+{1\over 2}G_{MN} \Lambda+{1\over 2}
 {\sqrt{-{\widehat G}}\over\sqrt{-G}}{\delta_M}^\mu{\delta_N}^\nu
 {\widehat G}_{\mu\nu}{\widetilde  f}\delta(z)=0~,
\end{equation}
where ${\widetilde f}\equiv {f/M_P^{D-2}}$. 

{}In the following we will be interested in solutions 
to the equations of motion following from the
action (\ref{action}) with the warped metric of the following form:
\begin{equation}\label{warped}
 ds_D^2=\exp(2A)\eta_{MN} dx^M dx^N~,
\end{equation}
where $\eta_{MN}$ is the flat $D$-dimensional Minkowski metric, and
the warp factor $A$, which is a function of $z$, is independent of the 
coordinates $x^\mu$. With this 
ans{\"a}tz, we have the following
equations of motion for $A$ (prime denotes derivative w.r.t. $z$):
\begin{eqnarray}
 \label{A'}
 &&(D-1)(D-2)(A^\prime)^2+ \Lambda \exp(2A)=0~,\\
 \label{A''}
 &&(D-2)\left[A^{\prime\prime}-(A^\prime)^2\right]
 =-{1\over 2} \exp(A) {\widetilde f}\delta(z)~.
\end{eqnarray}
This system of equations has a
solution if we fine-tune the brane tension to the bulk vacuum energy density
as follows:
\begin{equation}\label{fine}
 \Lambda=-{1\over 16}{{D-1}\over{D-2}}{\widetilde f}^2~.
\end{equation}
Assuming that the brane tension is positive, 
the solution is then given by
\begin{equation}\label{solution}
 A(z)=-\ln\left[{|z|\over\Delta}+1\right]~,
\end{equation}
where $\Delta\equiv {4(D-2)/{\widetilde f}}$, and we have chosen the 
integration constant such that $A(0)=0$.

{}The volume of the transverse dimension, which is given by
$\int dz \exp(DA)$, is finite in the above solution, and so is the
normalization of the $(D-1)$-dimensional graviton zero mode wave-function,
which is given by $\int dz \exp[(D-2)A]$. That is, gravity is localized on 
the brane, and at long distances we expect that gravity is
$(D-1)$-dimensional. In particular, at long distances we expect 
$(D-1)$-dimensional Newton's law
to be valid for matter sources localized on the brane, and the number of 
the graviton degrees of freedom which couple to the corresponding conserved 
energy-momentum tensor on the brane should be $(D-1)(D-4)/2$ (which for 
$D=5$ is 2). On the other hand, at
short distances we expect that gravity is $D$-dimensional. In particular,  
at short distances we expect the $D$-dimensional Newton's law to take over. 
As to the tensor structure of graviton propagator, consistency tells us that
we should also expect it to be that of the $D$-dimensional massless graviton
propagator. The number of the graviton degrees of freedom which couple to the
conserved energy-momentum tensor on the brane should then be 
$(D-2)(D-3)/2=(D-1)(D-4)/2+1$ (which for $D=5$ is 3)\footnote{Note that the
other $D-3=D(D-3)/2-(D-2)(D-3)/2$ degrees of freedom (that is, those 
corresponding to the graviphoton) do not couple to the conserved 
energy-momentum tensor on the brane.}. 

\section{Brane World Gravity}

{}To understand how such a smooth crossover between the number of relevant
degrees of freedom occurs in this model,  
let us study small fluctuations around the solution:
\begin{equation}\label{fluctu}
 G_{MN}=\exp(2A)\left[\eta_{MN}+{\widetilde h}_{MN}\right]~,
\end{equation}
where for convenience reasons we have chosen to work with 
${\widetilde h}_{MN}$
instead of metric fluctuations $h_{MN}=\exp(2A){\widetilde h}_{MN}$. 

{}To proceed further, we need equations of motion for ${\widetilde h}_{MN}$. 
Let us assume that we have matter localized on the brane, and 
let the corresponding conserved energy-momentum tensor be $T_{\mu\nu}$:
\begin{equation}\label{conserved}
 \partial^\mu T_{\mu\nu}=0~.
\end{equation}
The graviton field ${\widetilde h}_{\mu\nu}$ couples to $T_{\mu\nu}$ via
the following term in the action (note that ${\widetilde h}_{\mu\nu}=
h_{\mu\nu}$ at $z=0$ as we have set $A(0)=0$):
\begin{equation}\label{int}
 S_{\rm {\small int}}={1\over 2} \int_\Sigma d^{D-1} x ~T_{\mu\nu}
 {\widetilde h}^{\mu\nu}~.
\end{equation} 
Next, starting 
from the action $S+S_{\rm{\small int}}$ we obtain the
following linearized equations of motion for ${\widetilde h}_{MN}$
(the capital Latin indices $M,N,\dots$ are lowered and raised with the
flat $D$-dimensional Minkowski metric $\eta_{MN}$ and its inverse):
\begin{eqnarray}
 &&\left\{\partial_S\partial^S {\widetilde h}_{MN} +\partial_M\partial_N
 {\widetilde h}-\partial_M \partial^S {\widetilde h}_{SN}-
 \partial_N \partial^S {\widetilde h}_{SM}-\eta_{MN}
 \left[\partial_S\partial^S {\widetilde h}-\partial^S\partial^R
 {\widetilde h}_{SR}\right]\right\}+\nonumber\\
 &&(D-2)A^\prime\left\{\left[\partial_S {\widetilde h}_{MN} -
 \partial_M {\widetilde h}_{NS}-\partial_N{\widetilde h}_{MS}\right] n^S
 +\eta_{MN}\left[2\partial^R {\widetilde h}_{RS} - \partial_S 
 {\widetilde h}\right] n^S\right\}+\nonumber\\
 \label{EOMh}
 &&(D-1)(D-2)(A^\prime)^2 
 \eta_{MN}{\widetilde h}_{SR}n^S n^R=
 -M_P^{2-D} {\widetilde T}_{MN}\delta(z)~,
\end{eqnarray}
where ${\widetilde h}\equiv {\widetilde h}_M^M$, 
and for notational convenience we have introduced a unit vector $n^M$
with the following components: $n^\mu=0$, $n^D=1$. The
components of ${\widetilde T}_{MN}$ are given by:  
\begin{eqnarray}
 &&{\widetilde T}_{\mu\nu}\equiv T_{\mu\nu} -{1\over 2}\eta_{\mu\nu}
 f\rho~,\\
 &&{\widetilde T}_{\mu\nu}=A_\mu f~,\\
 &&{\widetilde T}_{DD}=0~,
\end{eqnarray}
Here and in the following we use the notations:
\begin{equation}
 H_{\mu\nu}\equiv{\widetilde h}_{\mu\nu}~,~~~
 A_\mu\equiv{\widetilde h}_{\mu D}~,~~~\rho\equiv {\widetilde h}_{DD}~. 
\end{equation}
Also, we will use the notation $H\equiv H_\mu^\mu$, and the Greek indices
$\mu,\nu,\dots$ are lowered and raised with the flat $(D-1)$-dimensional 
Minkowski metric $\eta_{\mu\nu}$ and its inverse.

{}The above equations of motion are invariant under certain gauge
transformations corresponding to unbroken diffeomorphisms.  
In terms of ${\widetilde h}_{MN}$ the full
$D$-dimensional diffeomorphisms 
\begin{equation}
 \delta h_{MN}=\nabla_M\xi_N+\nabla_N\xi_M
\end{equation}
are given by the following gauge 
transformations (here we use $\xi_M\equiv \exp(2A){\widetilde \xi}_M$):
\begin{equation}\label{gauge}
 \delta{\widetilde h}_{MN}=\partial_M {\widetilde\xi}_N+
 \partial_N{\widetilde\xi}_M+2A^\prime\eta_{MN}{\widetilde \xi}_S n^S~. 
\end{equation}
Note, however, that the presence of the brane
(whose tension is non-vanishing)
breaks the full $D$-dimensional diffeomorphism invariance (\ref{gauge}) to
a smaller subset of gauge transformations. To deduce the unbroken gauge
transformations, let us first rewrite the equations of motion (\ref{EOMh})
in terms of the components:
\begin{eqnarray}\label{EOM1o}
 &&\left\{\partial_\sigma\partial^\sigma 
 H_{\mu\nu} +\partial_\mu\partial_\nu
 H-\partial_\mu \partial^\sigma H_{\sigma\nu}-
 \partial_\nu \partial^\sigma H_{\sigma\mu}-\eta_{\mu\nu}
 \left[\partial_\sigma\partial^\sigma H-\partial^\sigma\partial^\rho
 H_{\sigma\rho}\right]\right\}+\nonumber\\
 &&\left\{H_{\mu\nu}^{\prime\prime}-\eta_{\mu\nu}H^{\prime\prime}+
 (D-2)A^\prime\left[H_{\mu\nu}^\prime-\eta_{\mu\nu}H^\prime\right]\right\}-
 \nonumber\\
 &&\left\{\partial_\mu A_\nu^\prime + \partial_\nu A_\mu^\prime -
 2\eta_{\mu\nu}\partial^\sigma A_\sigma^\prime+ (D-2)A^\prime
 \left[\partial_\mu A_\nu +\partial_\nu A_\mu
 -2\eta_{\mu\nu}\partial^\sigma
 A_\sigma\right]\right\}\nonumber+\\
 &&\left\{\partial_\mu\partial_\nu\rho-\eta_{\mu\nu}
 \partial_\sigma\partial^\sigma 
 \rho+\eta_{\mu\nu}\left[(D-2)A^\prime\rho^\prime
 +(D-1)(D-2)(A^\prime)^2\rho\right]\right\}= -M_P^{2-D} 
{\widetilde T}_{\mu\nu}
 \delta(z)~,\\ 
 \label{EOM2o} 
 &&\left[\partial^\mu H_{\mu\nu}-\partial_\nu H\right]^\prime 
 -\partial^\mu F_{\mu\nu}+(D-2)A^\prime 
 \partial_\nu\rho=M_P^{2-D}{\widetilde T}_{\mu D}\delta(z)~,\\
 \label{EOM3o}
 &&-\left[\partial^\mu\partial^\nu H_{\mu\nu}-
\partial^\mu\partial_\mu H\right]
 +(D-2) A^\prime \left[H^\prime-2\partial^\sigma A_\sigma\right]
 -(D-1)(D-2)(A^\prime)^2\rho=0~,
\end{eqnarray}
where $F_{\mu\nu}\equiv \partial_\mu A_\nu-\partial_\nu A_\mu$ is the
$U(1)$ field strength for the graviphoton.

{}In terms of the component fields $H_{\mu\nu}$, $A_\mu$ and $\rho$, the
full $D$-dimensional diffeomorphisms read:
\begin{eqnarray}\label{diff1}
 &&\delta H_{\mu\nu}=\partial_\mu{\widetilde\xi}_\nu+\partial_\nu{\widetilde
 \xi}_\mu+2\eta_{\mu\nu}A^\prime\omega~,\\
 \label{diff2}
 &&\delta A_\mu={\widetilde\xi_\mu}^\prime +\partial_\mu \omega~,\\
 \label{diff3}
 &&\delta\rho=2\omega^\prime+2A^\prime\omega~,
\end{eqnarray}
where $\omega\equiv{\widetilde\xi}_D$.
It is not difficult to check that (\ref{EOM3o}) is invariant under these
gauge transformations. On the other hand, the invariance of (\ref{EOM1o}) 
requires $\left[\omega\delta(z)\right]^\prime=0$. This then implies that 
$\omega(z=0)=0$. Finally, the invariance of (\ref{EOM2o}) can be seen to
imply ${\widetilde \xi}_\mu^\prime\delta(z)=0$, which implies 
${\widetilde \xi}_\mu^\prime(z=0)=0$.
Thus, the unbroken diffeomorphisms are given by (\ref{diff1}),
(\ref{diff2}) and (\ref{diff3}) subject to the following conditions:
\begin{eqnarray}
 &&\label{diffcond1}
 \omega(z=0)=0~,\\
 \label{diffcond2}
 &&{\widetilde \xi}_\mu^\prime(z=0)=0~.
\end{eqnarray}
Note that it is the presence of a non-zero tension brane that is
responsible for this reduction of gauge symmetry in the system.
The condition (\ref{diffcond1}) can be intuitively
understood by noting that $A^\prime$
is discontinuous at $z=0$, so that $\omega$ must vanish at $z=0$ or else
the gauge transformation (\ref{diff1}) for $H_{\mu\nu}$ would be 
discontinuous. On the other hand, (\ref{diffcond2}) becomes evident upon
examination of (\ref{EOM2o}). The r.h.s. of this equation, which is
given by $M_P^{2-D}{\widetilde T}_{\mu D}\delta(z)=A_\mu {\widetilde f}
\delta(z)$, is nothing but the mass term for the graviphoton on the brane,
so that the corresponding gauge transformation on the brane should be
vanishing. 

{}Before we turn to solving the above system of equations, the following
observation is in order. Note that the l.h.s. of (\ref{EOM2o}) does not 
contain terms with second derivatives w.r.t. $z$. This then implies that, to
have a consistent solution, we must have
\begin{equation}
 A_\mu(z=0)=0~.
\end{equation}
In fact, this is consistent with the fact that the graviphoton does not
couple to the conserved energy-momentum tensor on the brane.

\subsection{Partial Gauge Fixing}

{}Note that a physical solution to the above system of 
equations must satisfy the physical boundary conditions, namely, that all
perturbations should decay to zero away from the brane. In particular, the
graviphoton field $A_\mu$ must satisfy this requirement as well. 
It is not difficult to see that, since it 
vanishes on the brane, to have $A_\mu\rightarrow 0$ at 
$z\rightarrow\pm\infty$,
we must assume that $A_\mu$ vanishes everywhere. Note however, that this 
actually is not an additional requirement. Indeed, we can always gauge $A_\mu$
away. Thus, it vanishes on the brane. On the other hand, away from the brane
(that is, at $z\not=0$) we can always use the unbroken diffeomorphisms
(\ref{diff2}) to set $A_\mu$ to zero. Thus, consider diffeomorphisms with 
$\omega\equiv 0$. We can then always choose ${\widetilde \xi}_\mu$ such that
the gauge transformed $A_\mu$ identically vanishes. In the following 
we will refer to this 
gauge fixing as partial gauge fixing (as opposed to the complete gauge fixing
which we will discuss in the next subsection).  

{}After the aforementioned partial gauge fixing, we
have the following equations of motion:
\begin{eqnarray}\label{EOM1}
 &&\left\{\partial_\sigma\partial^\sigma 
 H_{\mu\nu} +\partial_\mu\partial_\nu
 H-\partial_\mu \partial^\sigma H_{\sigma\nu}-
 \partial_\nu \partial^\sigma H_{\sigma\mu}-\eta_{\mu\nu}
 \left[\partial_\sigma\partial^\sigma H-\partial^\sigma\partial^\rho
 H_{\sigma\rho}\right]\right\}+\nonumber\\
 &&\left\{H_{\mu\nu}^{\prime\prime}-\eta_{\mu\nu}H^{\prime\prime}+
 (D-2)A^\prime\left[H_{\mu\nu}^\prime-\eta_{\mu\nu}H^\prime\right]\right\}+
 \nonumber\\
 &&\left\{\partial_\mu\partial_\nu\rho-\eta_{\mu\nu}
 \partial_\sigma\partial^\sigma 
 \rho+\eta_{\mu\nu}\left[(D-2)A^\prime\rho^\prime
 +(D-1)(D-2)(A^\prime)^2\rho 
 \right]\right\}= -M_P^{2-D} {\widetilde T}_{\mu\nu}
 \delta(z)~,\\ 
 \label{EOM2} 
 &&\left[\partial^\mu H_{\mu\nu}-\partial_\nu H\right]^\prime +(D-2)A^\prime 
 \partial_\nu\rho=0~,\\
 \label{EOM3}
 &&-\left[\partial^\mu\partial^\nu H_{\mu\nu}-\partial^\mu\partial_\mu 
 H\right]
 +(D-2) A^\prime H^\prime -(D-1)(D-2)(A^\prime)^2\rho=0~.
\end{eqnarray}
To solve this system of equations, it is convenient to
perform the Fourier transform for the coordinates $x^\mu$ (we will denote the
corresponding momenta via $p^\mu$), and 
Wick rotate to the Euclidean space (where the propagator is unique). 
The equations of motion for the Fourier transformed
quantities read:
\begin{eqnarray}\label{EOM1F}
 &&-\left\{p^2 
 H_{\mu\nu} +p_\mu p_\nu
 H-p_\mu p^\sigma H_{\sigma\nu}-
 p_\nu p^\sigma H_{\sigma\mu}-\eta_{\mu\nu}
 \left[p^2 H-p^\sigma p^\rho
 H_{\sigma\rho}\right]\right\}+\nonumber\\
 &&\left\{H_{\mu\nu}^{\prime\prime}-\eta_{\mu\nu}H^{\prime\prime}+
 (D-2)A^\prime\left[H_{\mu\nu}^\prime-\eta_{\mu\nu}H^\prime\right]\right\}+
 \nonumber\\
 &&\left\{-p_\mu p_\nu\rho
 +\eta_{\mu\nu}\left[(D-2)A^\prime\rho^\prime +\rho \left(p^2
 +(D-1)(D-2)(A^\prime)^2\right)\right]\right\}= 
 -M_P^{2-D} {\widetilde T}_{\mu\nu}(p)
 \delta(z)~,\\ 
 \label{EOM2F} 
 &&\left[p^\mu H_{\mu\nu}-p_\nu H\right]^\prime +(D-2)A^\prime 
 p_\nu\rho=0~,\\
 \label{EOM3F}
 &&\left[p^\mu p^\nu H_{\mu\nu}-p^2 H\right]
 +(D-2) A^\prime H^\prime -(D-1)(D-2)(A^\prime)^2\rho=0~.
\end{eqnarray} 
Let us assume that $T(p)\equiv T_\mu^\mu(p)\not=0$.
Then the most general tensor structure for the fields $H_{\mu\nu}$ and
$\rho$ can be parametrized in terms of four functions $a,b,c,d$ as follows:
\begin{eqnarray}\label{d}
 &&\rho=M_P^{2-D}~d ~T(p)~,\\
 \label{abc}
 &&H_{\mu\nu}=M_P^{2-D}\left\{a~T_{\mu\nu}(p) 
 +\left[b~\eta_{\mu\nu}+c~p_\mu p_\nu
 \right]T(p)\right\}~.
\end{eqnarray}
Plugging this back into the equations of motion, we obtain five equations
for four unknowns $a,b,c,d$. However,
one of these equations is identically satisfied once we take into account 
the other four (as well as the on-shell expression for $A$). 
After some straightforward computations we obtain the following
system of four independent equations:
\begin{eqnarray}\label{eomQ1}
 &&a^{\prime\prime}+(D-2)A^\prime a^\prime-p^2 a=-\delta(z)~,\\
 \label{eomQ2}
 &&(D-2)A^\prime d=a^\prime+(D-2)b^\prime~,\\
 \label{eomQ3}
 &&A^\prime\left[(D-2)p^2c^\prime-a^\prime\right]=p^2
 \left[a+(D-2)b\right]~,\\
 \label{eomQ4}
 &&a+(D-3)b-c^{\prime\prime}-(D-2)A^\prime c^\prime+d=0~.
\end{eqnarray}
Here we note that (\ref{eomQ1}) must be solved subject to the boundary 
conditions $a(z\rightarrow\pm\infty)=0$ (for $p^2>0$).

{}Note that (\ref{eomQ4}) contains the second derivative of $c$ but
no $\delta$-function term. This then implies that $c^\prime$ must be
continuous at $z=0$. However, since $A^\prime$ is discontinuous at $z=0$,
the only way that we can satisfy (\ref{eomQ3}) is then to assume that
$c^\prime$ vanishes at $z=0$. Taking into account that at 
$z\rightarrow\pm\infty$ $c$ must go to zero, we can assume that
$c$ identically vanishes. Then we have only three unknowns left, namely, 
$a,b,d$, but we still have four equations. However, one of these equations, 
namely, (\ref{eomQ2}), is automatically satisfied once we take into account 
the other three equations (as well as the on-shell expression for $A$).
In terms of the solution to (\ref{eomQ1}) (which can be expressed via
Bessel functions \cite{RS}), we thus have the following expression for $b$
\begin{equation}    
 b=-{1\over{D-2}}\left[a+{1\over p^2}A^\prime a^\prime\right]~.
\end{equation}
Note that $d=-a-(D-3)b$.

{}Using the above results we obtain: 
\begin{eqnarray}\label{main}
 && H_{\mu\nu}(p,z)=M_P^{2-D} a(p,z) \left\{T_{\mu\nu}(p)-{1\over {D-2}}
 \left[1+
 {1\over p^2}A^\prime(z) \left(\ln[a(p,z)]\right)^\prime\right]
 \eta_{\mu\nu} T(p)
 \right\}~,\\
 \label{main1}
 && \rho(p,z)=-M_P^{2-D}{a(p,z)\over{D-2}} \left\{1-{{D-3}\over p^2}
 A^\prime(z)\left(\ln[a(p,z)]\right)^\prime\right\} T(p)~.
\end{eqnarray}
This solution gives
the form of (linearized) gravitational interactions between
matter sources localized on the brane including the tensor structure of the
corresponding graviton propagator. In fact, this is a physical
solution - it satisfies the physical boundary conditions at 
$z\rightarrow\pm\infty$, where both $H_{\mu\nu}(p,z)$ and 
$\rho(p,z)$ decay to zero as they should. In fact, in the following we will
argue that the leading ultra-violet behavior of $H_{\mu\nu}$ and $\rho$
on the brane 
following from the above solution is independent of the gauge choice.

\subsection{Complete Gauge Fixing}

{}Recall that above we set $c$ to zero, which left us with three unknowns 
$a,b,d$ yet four equations. However, as we mentioned above, only three out 
of these four equations are independent. This gives us a hint that, before 
we set $c$ to zero, there was
still some residual gauge symmetry left in the system. This, in fact, is 
indeed
the case. The simplest way to see this is to note that we can actually gauge 
both $A_\mu$ and $\rho$ away. 
Indeed, even though $\omega$ must vanish at $z=0$,
$\omega^\prime$ need not. It is then not difficult to see that we can 
simultaneously gauge $A_\mu$ and $\rho$ away.

{}One way to do this is to actually start from the solution (\ref{main}) and
(\ref{main1}), and find a gauge transformation that gauges $\rho$ away but 
does not affect $A_\mu$, which is vanishing. The corresponding gauge
parameters are given by ($\Delta$ was defined after (\ref{solution})):
\begin{eqnarray}
 &&\omega={M_P^{2-D}\over 4(D-2)\Delta} {T(p)\over p^2}
 \left[2a^\prime\Delta+{\rm sign}(z)
\left(|z|+\Delta\right)\right]~,\\
 &&{\widetilde \xi}_\mu={M_P^{2-D}\over 8(D-2)\Delta}{{i p_\mu T(p)}\over
 p^2}\left[4a\Delta+z^2+2\Delta|z|\right]~,
\end{eqnarray}
where we are working with the Fourier transformed quantities. 
Note that these gauge parameters formally 
satisfy the conditions (\ref{diffcond1})
and (\ref{diffcond2}). Here we note that ${\widetilde \xi}_\mu$ is given by
the above expression up to an arbitrary additive contribution independent of
$z$, which can be absorbed into $(D-1)$-dimensional diffeomorphisms (for which
$\omega\equiv 0$, and ${\widetilde \xi}_\mu$ are independent of $z$).

{}Next, the gauge transformed field
$H_{\mu\nu}$ is given by:
\begin{eqnarray}
 &&H_{\mu\nu}M_P^{D-2}=a\left[T_{\mu\nu}(p)-{1\over{D-2}}\eta_{\mu\nu}T(p)
 \right]-{1\over 2(D-2)\Delta}\eta_{\mu\nu}{T(p)\over p^2}+\nonumber\\ 
 \label{inconsistent}
 &&{1\over 4(D-2)\Delta}{{p_\mu p_\nu T(p)}\over p^2}\left[4a\Delta 
 +z^2+2\Delta |z|\right]~.
\end{eqnarray}
This expression has some peculiar properties. 
First, the last term diverges
as $z\rightarrow\pm\infty$, so that $H_{\mu\nu}$ given by this expression is
no longer a small perturbation around the background. Moreover, even though
this term is proportional to $p_\mu p_\nu$, 
a probe bulk matter source, for which $p^\mu T^{\rm{\small bulk}}_{\mu\nu}(p)$
need not vanish, 
will feel this field away from the brane. Second, even if we consider
only bulk matter sources with $p^\mu T^{\rm{\small bulk}}_{\mu\nu}(p)=0$,
so that the last term in (\ref{inconsistent}) cannot be measured, the second
term in (\ref{inconsistent}) is still felt by such a probe bulk matter source
(unless its energy-momentum tensor is traceless). This term, however, is
independent of $z$ (while the first term in (\ref{inconsistent}) decays to 
zero
at $z\rightarrow\pm\infty$), so there is a non-vanishing perturbation 
even infinitely far away form the brane. That is, the above gauge transformed
solution does not satisfy the physical boundary conditions at 
$z\rightarrow\pm\infty$. 

{}Another way of arriving at the above result is to use the complete gauge 
fixing to begin with, that is, to set $A_\mu$ as well as $\rho$ to zero 
before 
solving the equations of motion. This amounts to setting $d$ to zero in
the system of equations (\ref{eomQ1}), (\ref{eomQ2}), (\ref{eomQ3}) and
(\ref{eomQ4}). Once again, we now have three unknowns $a,b,c$ and four 
equations. However, as before, only three of these equations are independent.
In fact, upon solving this system of equations for $a,b,c$, we obtain 
precisely
(\ref{inconsistent}) (up to terms which can be gauged away using the 
aforementioned $(D-1)$-dimensional diffeomorphisms). 

{}Thus, as we see, the scalar degree of freedom $\rho$ cannot be gauged away
in the sense that, if we do gauge it away, the corresponding gauge 
transformed solution is not physical. In particular, it does not satisfy the
physical boundary conditions at $z\rightarrow\pm\infty$. 
However, above we saw that we cannot gauge $\rho$ away everywhere, but
we could ask whether $\rho$ can be gauged away on the brane (while it is 
non-vanishing in the bulk) and still satisfy the physical boundary conditions.

{}The answer to this question is positive, but there is an important point 
arising in such gauge fixing that we would like to discuss here.
Thus, consider a gauge transformation with the following properties:
\begin{equation}\label{xiCGF}
 {\widetilde \xi}_\mu(z)=ip_\mu\int_0^z dz_1 ~\omega(z_1)~,
\end{equation}
where $\omega(z)$ satisfies the following conditions:
\begin{equation}\label{gaugeBB}
 \omega(0)=0~,~~~\omega^\prime(0)={1\over 2(D-2)} {T(p)\over p^2}
 \left[p^2 a(p,0)-{{D-3}\over 2\Delta}\right]~,
\end{equation}
and $\omega(z)$ goes to zero at $z\rightarrow\pm\infty$ fast enough so that
${\widetilde \xi}_\mu(z)$ also goes to zero. Moreover, let us require that
at any finite $z$ none of the quantities $\omega(z)$, $\omega^\prime(z)$ and
${\widetilde \xi}_\mu (z)$ have poles for any values of $p^2$ even after we 
rotate back to the Minkowski space. In fact, it is non-trivial that this last
condition can be satisfied. Thus, consider ${\widetilde \xi}_\mu (z)$ 
infinitesimally close to the brane. From (\ref{xiCGF}) we have   
\begin{equation}
 {\widetilde \xi}_\mu(z)={{ip_\mu}\over 2}\omega^\prime(0) z^2+{\cal O}(z^3)~.
\end{equation}
On the other hand, for small $p^2$ we have $a(p,0)=(D-3)/2\Delta p^2+
{\cal O}(\Delta)$, so that $\omega^\prime(0)$ does not have a pole at $p^2=0$,
nor does ${\widetilde \xi}_\mu (z)$ infinitesimally near the brane. 
The above gauge transformation then introduces no dangerous poles in 
$H_{\mu\nu}$ or $\rho$. Moreover, it is not difficult to see that the gauge
transformed graviphoton still vanishes everywhere. The gauge transformed 
graviscalar $\rho$ now vanishes on the brane, but it is non-vanishing outside 
of the brane. Finally, the above gauge transformation does not affect the
graviton components on the brane as $\omega(0)=0$ and 
${\widetilde\xi}_\mu(0)=0$, and we have
\begin{equation}\label{mainCGF}
 H_{\mu\nu}(p,0)M_P^{D-2}=a(p,0)
 \left[T_{\mu\nu}(p)-{1\over{D-2}}\eta_{\mu\nu}T(p)
 \right]-{1\over 2(D-2)\Delta}\eta_{\mu\nu}{T(p)\over p^2}~.
\end{equation} 
Here we note that this result is the same as that obtained using the 
``brane bending''
procedure of \cite{GT,GKR}\footnote{Some issues in the ``brane bending''
procedure were discussed in \cite{Mueck}.}. In fact, the gauge transformation
with the properties (\ref{xiCGF}) and (\ref{gaugeBB}) precisely corresponds to
the ``brane bending'' procedure, where the graviscalar component vanishes on
the brane due to the fact that the ``brane bending'' brings us to the Gaussian
normal coordinates w.r.t. the brane, where the $(DD)$ component of the metric
perturbation vanishes \cite{GT,GKR}.

{}There is, however, a subtlety arising in using the above gauge fixing which
sets $\rho$ to zero. From the above discussion it is clear that there is no
subtlety as far as the infra-red behavior of the graviton propagator is 
concerned. The subtlety, however, does arise in the {\em ultra-violet}. Thus,
consider the variations due to the above gauge transformation in 
$H_{\mu\nu}$ infinitesimally near the brane:
\begin{eqnarray}
 \delta H_{\mu\nu}(z)= &&-ip_\mu{\widetilde \xi}_\nu-ip_\nu{\widetilde\xi}_\mu
 +2\eta_{\mu\nu}A^\prime\omega=\nonumber\\
 \label{deltaH}
 &&p_\mu p_\nu\omega^\prime(0) z^2\left[1+{\cal O}
 (z)\right] -\eta_{\mu\nu}
 {2\over \Delta}\omega^\prime(0)|z|\left[1+{\cal O}
 (z)\right]~.
\end{eqnarray} 
It is not difficult to see that at large $p^2$, that is, for $p^2\Delta^2\gg1$,
we have $a(p,0)\approx 1/2p$, where $p\equiv \sqrt{p^2}$. This then implies 
that at large $p^2$ we have
\begin{equation}
 \omega^\prime(0)\approx{1 \over {2(D-2)}}{T(p)\over {2p}}~.
\end{equation}
Thus, as we see, in (\ref{deltaH}) 
the term proportional to $p_\mu p_\nu$ has the following 
momentum structure at large $p^2$:
\begin{equation}
 {{p_\mu p_\nu}\over p}~,
\end{equation}
so that the graviton propagator blows up in the ultra-violet infinitesimally 
near the brane. Such a behavior is not acceptable for the graviton propagator,
in particular, the corresponding perturbation is no longer small for large
$p^2$. This implies that the diffeomorphisms required for setting $\rho$
to zero on the brane actually are not allowed. 

{}Here, however, we would like to discuss the diffeomorphisms that do not
lead to ultra-violet inconsistencies of the aforementioned type. Thus, 
consider diffeomorphisms where we still have (\ref{xiCGF}) as well as 
$\omega(0)=0$, but $\omega^\prime(0)$ goes to zero at large $p^2$ as
\begin{equation}
 \omega^\prime(0)\sim {1\over p^\alpha}~.
\end{equation}
(We are assuming that $\omega(z)$ is such that this gauge transformation does
not introduce any infra-red problems, in particular, it does not introduce any
dangerous poles.) For the gauge transformed $H_{\mu\nu}$ not to blow up at
large $p^2$, we must then assume that $\alpha\geq 2$. Note that, if this 
condition is satisfied, the gauge transformation in $\rho(z=0)$ only 
introduces terms which are subleading compared with the leading behavior
of $\rho(z=0)$ given by (\ref{main1}) at large $p^2$. 
That is, the allowed 
diffeomorphisms are such that they introduce only subleading terms into 
$\rho(z=0)$ at large $p^2$. In fact, the corresponding gauge parameter
$\omega^\prime(0)$ vanishes in the ultra-violet limit. This fact will become
important when we discuss quantum corrections on the brane.

\subsection{Long and Short Distance Behavior}

{}Let us now use (\ref{main}) to obtain the tensor and momentum structures of
the graviton propagator at small and large $p$. At large distance scales $r\gg 
\Delta$, that is, at small 
momenta $p\ll 1/\Delta$, we have 
\begin{equation}\label{a0}
 a(p,z=0)={1\over {Lp^2}}+\dots~,
\end{equation} 
where the ellipses stand for subleading corrections, and
\begin{equation}
 L\equiv {{\widehat M}_P^{D-3} \over{M_P^{D-2}}}=\int dz \exp[(D-2)A]=
 {2\over{D-3}} \Delta~.
\end{equation}
Here ${\widehat M}_P$ is the $(D-1)$-dimensional (reduced) Planck scale
corresponding to the $(D-1)$-dimensional Newton's constant which determines
the strength of gravitational interactions mediated by the localized graviton
zero mode. On the other hand, $A^\prime(0\pm)=\mp {1/\Delta}$, and 
$a^\prime(p,z=0\pm)=\mp {1/2}$ (the latter can be seen from (\ref{eomQ1})).
We therefore obtain that at small momenta
\begin{equation}
 H_{\mu\nu}(p,z=0) \approx {\widehat M}_P^{3-D}{1\over p^2} 
 \left[T_{\mu\nu}(p)-{1\over{D-3}}
 \eta_{\mu\nu} T(p)\right]~.
\end{equation}
Both the momentum and tensor structures in this expression 
for the gravitational field on the brane are 
$(D-1)$-dimensional. Thus, at large distances we have the $(D-1)$-dimensional 
$1/r^{D-4}$ Newton's law, and the number of degrees of freedom which couple to
$T_{\mu\nu}$ (which can be read off the coefficient $-1/(D-3)$ in front of the
$\eta_{\mu\nu} T(p)$ term) is $(D-1)(D-4)/2$. In particular, note that 
the coupling of the scalar degree of freedom $\rho\equiv{\widetilde h}_{DD}$ 
to brane matter is suppressed by an extra factor of order 
$p^2\Delta^2$ compared with that of the graviton $H_{\mu\nu}$.
This can be seen from the fact that at $p^2\rightarrow 0$ we have
$d(p,0)=-a(p,0)-(D-3)b(p,0)\rightarrow{\rm const.}\times\Delta$ (this follows
from the fact that the next-to-leading correction in (\ref{a0}) is 
${\cal O}(\Delta)$). This then implies that
\begin{equation}\label{longrho}
 \rho(p,z=0)\approx {\rm const.}\times {\widehat M}_P^{3-D}\Delta^2 T(p)~. 
\end{equation}
This, in particular, implies that at long distances we indeed have 
$(D-1)$-dimensional gravity. In fact, let us note that the long distance 
behavior of $\rho(p,z=0)$ given by (\ref{longrho}) is gauge dependent, and
it is not difficult to see that we can always go to a gauge where the 
coupling of $\rho(p,z=0)$ to the brane matter vanishes as $p^2\rightarrow 0$. 

{}Next, let us see what happens at small distance scales $r\ll\Delta$, that is,
at large momenta $p\gg1/\Delta$. We now have 
\begin{equation}\label{largea}
 a(p,z)\approx{1\over 2p}\exp(-p|z|)~,
\end{equation}
so that we obtain 
\begin{equation}\label{shortH}
 H_{\mu\nu}(p,z=0)\approx M_P^{2-D}{1\over 2p}
 \left[T_{\mu\nu}(p)-{1\over {D-2}}\eta_{\mu\nu}T(p)\right]~.
\end{equation}
Both the momentum and tensor structures in this expression 
for the gravitational field on the brane are 
$D$-dimensional. Thus, at small distances we have the $D$-dimensional 
$1/r^{D-3}$ Newton's law, and the number of degrees of freedom which couple to
$T_{\mu\nu}$ (which can be read off the coefficient $-1/(D-2)$ in front of the
$\eta_{\mu\nu} T(p)$ term) is $(D-2)(D-3)/2=(D-1)(D-4)/2+1$. The extra degree
of freedom is precisely the scalar $\rho$ which no longer decouples at large
momenta. This can be seen from the fact that at large $p$ we have 
$d(p,0)\approx -a(p,0)/(D-2)$, and
\begin{equation}\label{shortrho}
 \rho(p,z=0)\approx M_P^{2-D}{T(p)\over 2(D-2)p}~.
\end{equation} 
Note that that (\ref{shortH}) and (\ref{shortrho}) are in complete agreement
with the corresponding expressions in the case of a tensionless brane embedded
in the Minkowski bulk \cite{DGP} as they should be. 
Indeed, at large $p$ the leading behavior of 
gravity in Randall-Sundrum brane world is expected to be the same as in
the limit where the bulk vacuum energy density $\Lambda$ as well as the brane
tension $f$ go to zero with the fine-tuning relation (\ref{fine}) fixed.
In this limit we precisely 
have a tensionless brane in the Minkowski bulk.
On the other hand, as we pointed out in the previous subsection, the allowed
gauge transformations do not affect the leading behavior of $\rho(p,z=0)$ 
at large $p$ given by (\ref{shortrho}). That is, the coupling of $\rho$ to 
the brane matter is non-vanishing at large $p$, and it cannot be gauged 
away using the allowed diffeomorphisms.

\section{Implications}

{}The fact that $\rho$ couples to the brane matter in the ultra-violet
has important implications. Thus, note that the brane world-volume theory is 
not conformal as long
as gravity is localized. This then implies that quantum corrections will 
generically generate various terms in the brane world-volume
for the fields that couple to the brane matter. In the Randall-Sundrum model
these fields are $H_{\mu\nu}$ and $\rho$. It is clear that the corresponding
terms in the brane world-volume action should respect the $(D-1)$-dimensional
diffeomorphisms on the brane
(for these diffeomorphisms $\omega\equiv 0$, and ${\widetilde
\xi}_\mu$ are independent of $z$). Let us confine our attention
to such terms 
that are quadratic in $H_{\mu\nu}$ and/or $\rho$, and contain at most two
derivatives w.r.t. $x^\mu$. 
Then the most general corrections of this type into the brane
world-volume action are given by the following terms:
\begin{eqnarray}\label{correctionH}
 &&S_H={\widehat M}_P^{D-3} C_1 \int_\Sigma d^{D-1} x \left\{
 {1\over 4}\left[\partial_\sigma H\partial^\sigma H-\partial_\sigma
 H_{\mu\nu}\partial^\sigma H^{\mu\nu}\right]+{1\over 2}
 \left[\partial^\mu H_{\mu\sigma}\partial_\nu H^{\nu\sigma}-\partial^\mu
 H_{\mu\nu}\partial^\nu H\right]\right\}~,\nonumber\\
 \label{correctionrho}
 &&S_\rho={\widehat M}_P^{D-3} \int_\Sigma d^{D-1} x\left\{C_2
 \left[\partial_\nu H-\partial^\mu
 H_{\mu\nu}\right]\partial^\nu\rho -C_3 m_\rho^2\rho^2 -
 C_4\partial_\mu\rho \partial^\mu\rho\right\}~.
\end{eqnarray}   
The terms combined into $S_H$ correspond to expanding the 
term\footnote{The fact that this term is generically generated by quantum 
effects on a brane was pointed out in \cite{DGP,DG}. The fact that it should
be included in the Randall-Sundrum model was pointed out in \cite{zura}.}
\begin{equation}
 {\widehat M}_P^{D-3} C_1 \int_\Sigma d^{D-1} x\sqrt{-{\widehat G}}{\widehat
 R}
\end{equation}
around the Minkowski background: ${\widehat G}_{\mu\nu}=\eta_{\mu\nu}+
H_{\mu\nu}$ (${\widehat R}$ denotes the $(D-1)$-dimensional Ricci scalar
constructed from the metric ${\widehat G}_{\mu\nu}$). Note that this term
renormalizes the $(D-1)$-dimensional Planck scale, and one can show that
it does not introduce any inconsistency as far as the coupling between bulk
gravity and brane matter is concerned. 

{}On the other hand, as we will point out in a moment, the terms appearing 
in (\ref{correctionrho}) are a bit more
harmful. The term with the coefficient $C_2$ 
corresponds to expanding the term 
\begin{equation}
 {\widehat M}_P^{D-3} C_2 \int_\Sigma d^{D-1} x ~\rho\sqrt{-{\widehat G}}
 {\widehat R} 
\end{equation}
around the Minkowski background. The term proportional to $C_3$ (that is, the 
mass term for $\rho$) corresponds to expanding the term
\begin{equation}
 -{\widehat M}_P^{D-3} C_3 \int_\Sigma d^{D-1} x ~\sqrt{-{\widehat G}}V(\rho) 
\end{equation}
around the Minkowski background. Here $V(\rho)$ is the scalar potential for
$\rho$, which is also expanded w.r.t. $\rho$. The first term in this latter
expansion is given by 
\begin{equation}
 -{\widehat M}_P^{D-3} C_3 V(0) \int_\Sigma d^{D-1} x\sqrt{-{\widehat G}}~.
\end{equation} 
This term renormalizes the brane tension. Then to preserve $(D-1)$-dimensional
Poincar{\'e} invariance on the brane we need to accordingly fine-tune the
bulk vacuum energy density. This fine-tuning is the usual fine tuning of 
the $(D-1)$-dimensional cosmological constant, and it will not concern us 
here. On the other hand, the term linear in $\rho$, which is given by
\begin{equation}
 -{\widehat M}_P^{D-3}
 C_3 V_\rho (0) \int_\Sigma d^{D-1} x ~\rho\sqrt{-{\widehat G}}~,
\end{equation}
deserves a more careful consideration. Indeed, if $V_\rho(0)\not=0$, 
this term corresponds to a
tadpole for $\rho$ when expanded around the classical vacuum. It is not 
difficult to see that such a tadpole would destabilize the background. The 
reason for this is that in the presence of such a tadpole the r.h.s. of the
$(DD)$ component of (\ref{EinsteinEOM}), that is, the r.h.s. of (\ref{A'})
now contains the delta-function source term, while its l.h.s. does not contain
terms with the second derivative w.r.t. $z$. This implies that $\rho=0$ 
does not correspond to a consistent background. On the other hand, if $V_\rho$
vanishes for some other $\rho=\rho_0$, we must then expand around this point
instead of $\rho=0$. We can then always bring the corresponding solution to the
form we have been using by appropriate rescalings. The important point here, 
however, is that {\em a priori} there is no guarantee that the scalar potential
for $\rho$ has an extremum. Thus, for instance, $V(\rho)$ could have a runaway
behavior. That is, in this model we {\em a priori} 
have to deal with a type of the {\em moduli} problem.
In the following we will, however, assume that (possibly after the 
aforementioned rescalings) we have $V_\rho(0)=0$. Finally, let us mention
that the term in (\ref{correctionrho}) proportional to $C_4$ is the kinetic
term for $\rho$.

{}In fact, there is one more term which will be relevant for the following
discussion. Thus, at the quantum level there might be generated a non-vanishing
coupling of $\rho$ to the brane matter analogous to that for the graviton 
$H_{\mu\nu}$. That is, instead of (\ref{int}) we can consider a more general
coupling  
\begin{equation}\label{intgen}
 S_{\rm {\small int}}={1\over 2} \int_\Sigma d^{D-1} x ~T_{MN}
 {\widetilde h}^{MN}~.
\end{equation}
Here we will assume that $T_{\mu D}=0$ as $A_\mu$ does not couple to the
conserved energy-momentum tensor $T_{\mu\nu}$ on the brane. However,
the coupling $T_{DD}$ generically need not vanish at the quantum level.

{}Here we note that the terms in (\ref{correctionrho}) as well as the
$T_{DD}$ coupling in (\ref{intgen}) are not invariant under the
diffeomorphisms given by (\ref{diff1}), (\ref{diff2}) and (\ref{diff3})
subject to the conditions (\ref{diffcond1}) and (\ref{diffcond2}) (while the
correction given by $S_H$ is) as in these diffeomorphisms
$\omega^\prime(0)$ need not vanish. However, as we pointed out
in the previous section, not all of these diffeomorphisms are allowed - 
$\omega^\prime(0)$ must vanish fast enough with $p^2\rightarrow
\infty$ or else the gauge 
transformed graviton propagator blows up near the brane in the ultra-violet.
That is, generically there is nothing preventing the terms
in (\ref{correctionrho}) as well as the $T_{DD}$ coupling in (\ref{intgen}) 
from being generated at the quantum level. In fact, the corresponding 
counterterms are precisely related to {\em ultra-violet} 
divergences in the theory.

{}The trouble with the terms appearing in (\ref{correctionrho}) as well as
the aforementioned coupling $T_{DD}$ is that they
give rise to a source term on the r.h.s. of (\ref{EOM3}), while its l.h.s. 
does not contain terms with the second derivative w.r.t. $z$.
It is then clear that, to have a consistent
solution to (\ref{EOM3}), the total source term on its r.h.s. should vanish.
This, however, generically implies that we need to fine-tune parameters 
$C_2,C_3,C_4$ as well as the coupling $T_{DD}$ to have a continuous solution
for $\rho$ and $H_{\mu\nu}$. Indeed, the values of $\rho$ and $H_{\mu\nu}$ 
at $z=0$ are completely determined in terms of $T_{\mu\nu}$. Note that this
fine-tuning is independent of the aforementioned fine-tuning required for
maintaining vanishing cosmological constant on the brane.   

{}Note that if we naively perform the ``brane bending'' procedure, in the
corresponding Gaussian normal coordinates $\rho$ does not couple to the brane
matter, so in this gauge it might seem that the aforementioned dangerous
terms are not generated. However, in the presence of (non-conformal) matter
on the brane, in this gauge the brane is no longer
straight but is bent w.r.t. the coordinate system that is straight w.r.t.
the AdS horizon \cite{GT,GKR}. {\em A priori} it is then unclear how to compute
quantum corrections due to the brane matter loops on such a brane, so
that the conclusion that the aforementioned terms are not generated in this 
gauge might be incorrect. In fact, had such a conclusion been correct, we would
have a puzzle - physics should certainly be independent of the gauge choice, 
and in the gauge where $\rho$ is non-vanishing on the brane the corresponding
terms are generically expected to be generated. Note, however, that the 
resolution of this issue appears to be quite simple if we take into account
our previous discussions - the gauge corresponding to the Gaussian normal 
coordinates appears to suffer from ultra-violet divergences in the graviton 
propagator, and is not suitable for discussing the aforementioned 
quantum corrections on the brane.

{}Thus, as we see, the fact that $\rho$ cannot be gauged away necessitates
fine-tuning at the quantum level to preserve consistent coupling between
bulk gravity and brane matter. A similar observation was recently made
in \cite{zura2} in a somewhat different context, namely, within the setup 
of \cite{zura1} where the extra dimension has infinite volume. Because of
the latter fact, however, consistent coupling between bulk gravity and
brane matter can be achieved in the setup of \cite{zura1} if the brane
world-volume theory is conformal \cite{zura2}.  

\section{Remarks}

{}Let us now comment on quantum corrections in the bulk. In particular, as
was pointed out in \cite{COSM}, in warped backgrounds with finite-volume 
non-compact extra dimensions one must be cautious about higher derivative
terms in the bulk action. Note that as long as $-\Lambda\ll
M_*^2$, where $M_*$ is the cut-off scale for higher derivative terms in the
bulk action, then contributions of such terms as far as the domain wall 
solution is concerned are under control \cite{COSM}. 
However, as was pointed out in \cite{COSM}, higher curvature terms in such
warped compactifications generically 
lead to delocalization of gravity. Thus,
inclusion of higher derivative terms of, say, the form
\begin{equation}
 \zeta\int d^Dx \sqrt{-G} R^k
\end{equation}
into the bulk action would produce terms of the form \cite{COSM}
\begin{equation}
 \zeta \int d^{D-1}x dz\exp[(D-2k)A]\sqrt{-{\widehat G}}{\widehat R}^k~.
\end{equation}
Assuming that $A$ goes to $-\infty$ at $z \rightarrow\pm\infty$, 
for large enough $k$ the factor $\exp[(D-2k)A]$ 
diverges, so that at the end of the day gravity is no longer localized.

{}A possible way around this difficulty might be that all the higher curvature
terms should come in ``topological'' combinations (corresponding to Euler
invariants such as the Gauss-Bonnet term \cite{Zwiebach,Zumino}) 
so that their presence does not
modify the $(D-1)$-dimensional propagator for the bulk graviton modes. That is,
even though such terms are multiplied by diverging powers of the warp factor,
they are still harmless. One could attempt to justify the fact that higher
curvature bulk terms must arise only in such combinations by the fact that
otherwise the bulk theory would be inconsistent to begin with due to the
presence of ghosts. However, it is not completely obvious whether it is
necessary to have only such combinations to preserve unitarity. Thus, in
a non-local theory such as string theory unitarity might be preserved,
even though at each higher derivative order there are non-unitary terms, due
to a non-trivial cancellation between an infinite tower of such terms.

{}We would like to end our discussion by pointing out that the aforementioned
difficulty with higher curvature terms does not arise in theories with 
infinite-volume non-compact extra dimensions 
\cite{GRS,CEH,DGP0,witten,DGP,zura,zura1,zura2,DG}. 
However, consistency of the coupling between bulk 
gravity and brane matter might give rise to additional constraints. As we 
have already mentioned, in some cases the brane world-volume theory must be
conformal. In such cases it would be interesting to understand if there is a
relation to \cite{BKV}. 

\acknowledgments

{}I would like to thank Gia Dvali, Gregory Gabadadze and especially Tom Taylor
for valuable discussions. Parts of this work were completed while I was
visiting at Harvard University, New York University and Northeastern 
University.
This work was supported in part by the National Science Foundation.
I would also like to thank Albert and Ribena Yu for financial support.

\end{document}